\documentclass[useAMS,usenatbib]{mn2e}
\usepackage{graphicx}
\usepackage{amsmath}
\usepackage{tabularx}

\usepackage{times}
\usepackage{amssymb,amsmath}

\def\lsim{ \lower .75ex\hbox{$\sim$} \llap{\raise .27ex \hbox{$<$}} }
\def\gsim{ \lower .75ex \hbox{$\sim$} \llap{\raise .27ex \hbox{$>$}} }

\newcommand{\bi}{\begin{itemize}}
\newcommand{\ei}{\end{itemize}}

\title[Recollimation, instabilities and extreme blazars] 
{Extreme blazars: the result of unstable recollimated jets?} 

\author[Tavecchio, Costa, Sciaccaluga]
{Fabrizio Tavecchio$^1$\thanks{E--mail: fabrizio.tavecchio@inaf.it, agnese.costa@inaf.it, alberto.sciaccaluga@inaf.it}, Agnese Costa$^{2,1}$, Alberto Sciaccaluga$^{3,1}$\\
$^1$INAF -- Osservatorio Astronomico di Brera, via E. Bianchi 46, I--23807 Merate, Italy\\
$^2$Università degli Studi dell'Insubria, via Ravasi 2, I--21100 Varese, Italy\\
$^3$Dipartimento di Fisica, Università degli studi di Genova, Via Dodecaneso 33, 16146, Italy\\
}

\begin{document}



\maketitle

\begin{abstract} 
Extreme BL Lacs (EHBL) form a subclass of blazars which challenge standard emission scenarios. In a recent study it has been argued that their peculiar properties can be explained if emitting electrons are accelerated in a series of oblique shocks induced by the recollimation of the relativistic jet. However, new 3D simulations of recollimated, weakly magnetized jets reveal that, in correspondence with the first recollimation shock, the flow develops a rapidly growing instability, becomes highly turbulent and decelerates, effectively hampering the formation of the multiple shock structure routinely observed in 2D simulations. Building on these new findings, we propose here a revised scenario for EHBL, in which the emission is produced by electrons accelerated at the recollimation shock and subsequently further energized through stochastic acceleration in the turbulent downstream flow. We apply a simple version of this scenario to the prototypical EHBL 1ES 0229-200, showing that the SED can be satisfactorily reproduced with standard values of the main physical parameters.
\end{abstract}


\section{Introduction}

Blazars are, among extragalactic sources, the most abundant emitters at high and very high gamma-ray energies. Their relativistic jets, closely oriented toward the Earth, are expected to offer suitable conditions for which particles (leptons, possibly hadrons) can be accelerated at relativistic energies and can copiously emit non-thermal radiation (e.g. Romero et al. 2017, Blandford et al. 2019). Their spectral energy distribution (SED) covers the entire electromagnetic spectrum, from radio waves to gamma-rays (e.g., Ghisellini et al. 2017). The characteristic SED shape, displaying two ``humps", is commonly interpreted as due to synchrotron and inverse Compton (IC) emission from relativistic electrons (Ghisellini et al. 1998; but see e.g., Boettcher et al. 2013 for hadronic models). 
The powerful and rapidly variable radiative output of blazar flags the existence of efficient processes able to accelerate the emitting particles up to ultra-relativistic energies. Extensively investigated acceleration processes include diffusive shock acceleration (DSA), stochastic acceleration by (relativistic) turbulence (SA) and  magnetic reconnection (e.g. Matthews et al. 2020 for a review). 

Among blazars, the so-called extreme highly peaked BL Lacs  (EHBL, Costamante et al. 2001; see Biteau at al. 2020 for a review and for the relevant literature) stand out for several peculiar properties. In particular, the extremely hard GeV-TeV continuum extending at least up to 10 TeV is challenging for standard emission models (e.g.,Tavecchio et al. 2009, Costamante et al. 2018). Moreover, at odds with the bulk of the blazar population, the high-energy emission appears to be substantially stable. The unusual phenomenology of EHBL has been explained invoking an hard electron distribution with a large minimum energy (e.g. Tavecchio et al. 2009), a maxwellian-like electron distribution (Lefa et al. 2011) a beam of high-energy hadrons (e.g. Essey \& Kusenko 2010), internal absorption (e.g. Aharonian et al. 2008) or emission from a large-scale jet (B{\"o}ttcher et al. 2008).

Recently Zech \& Lemoine (2021) have reported a comprehensive analysis of the phenomenology of EHBLs in the light of recent results on DSA obtained through state-of-the-art simulations. The modeling of the SED convincingly show that, in the leptonic framework, the magnetic field of these sources is quite low  (few mG, e.g. Tavecchio et al. 2010, Costamante et al. 2018). In this condition DSA is expected to work efficiently (e.g. Sironi et al. 2015), while an important role for reconnection or relativistic magnetic turbulence can be confidently excluded.  Zech \& Lemoine (2021) analyzed several possible scenarios and concluded that the most likely possibility is that particles are energized by recollimation shocks expected to form as a result of the recollimation of the jet by external gas (e.g. Komissarov \& Falle 1998). While the particle energy distribution expected for one shock can reproduce the SED of some of the less extreme EHBL, a single shock is likely unable to provide an electron population with the properties (minimum energy, slope) required to reproduce the SED of the EHBL with the hardest high-energy spectra (among which the prototype 1ES 0229+200). Zech \& Lemoine (2021) argue that the observed emission properties of EHBL can be reproduced provided that particles undergo more than one acceleration step, by e.g. multiple shocks, naturally expected (as routinely confirmed by 2D numerical simulations) after the initial recollimation (e.g., Fichet de Clairfontaine G. et al. 2021). 

The phenomenology of jet recollimation and recollimation shocks has been commonly explored by means of 2D simulations (e.g. Mizuno et al. 2015, Bodo \& Tavecchio 2018, Fichet de Clairfontaine et al. 2021). Recent extensions to 3D (Gourgouliatos \& Komissarov 2018, Matsumoto et al. 2021) somewhat unexpectedly revealed that for hydrodynamical jets the recollimation trigger a strong instability (likely close to the classical centrifugal instability affecting rotating fluids, Komissarov et al. 2019) which rapidly grows, injects strong turbulence in the flow and eventually decelerates and disrupts the jet. Magnetic fields of sufficient intensity are able to damp the instability and stabilize the flow (Matsumoto et al. 2021). However, for low magnetization (such as that characterizing the jet of extreme blazars) the behavior of the instability is similar to the hydrodynamical case. These results forces to reconsider the scenario depicted by Zech \& Lemoine (2021). In particular, since the simulations clearly show that the flow becomes strongly turbulent after the recollimation shock, it is tempting to explore the possibility that particles energized at the shock can be further accelerated by the turbulence, with an effective interplay between DSA and SA (see e.g. Kundu et al. 2021). 

In this paper we intend to provide the results of a first exploration of the scenario sketched above, presenting a simple model able to reproduce the SED of 1ES 0229+200. In Sect. 2 we set the stage , in Sect. 3 we describe the model and in Sect. 4 we discuss the results. 

\section{A scenario for the extreme BL Lac objects}

\subsection{The framework}

\begin{figure}
\includegraphics[width=0.47\textwidth]{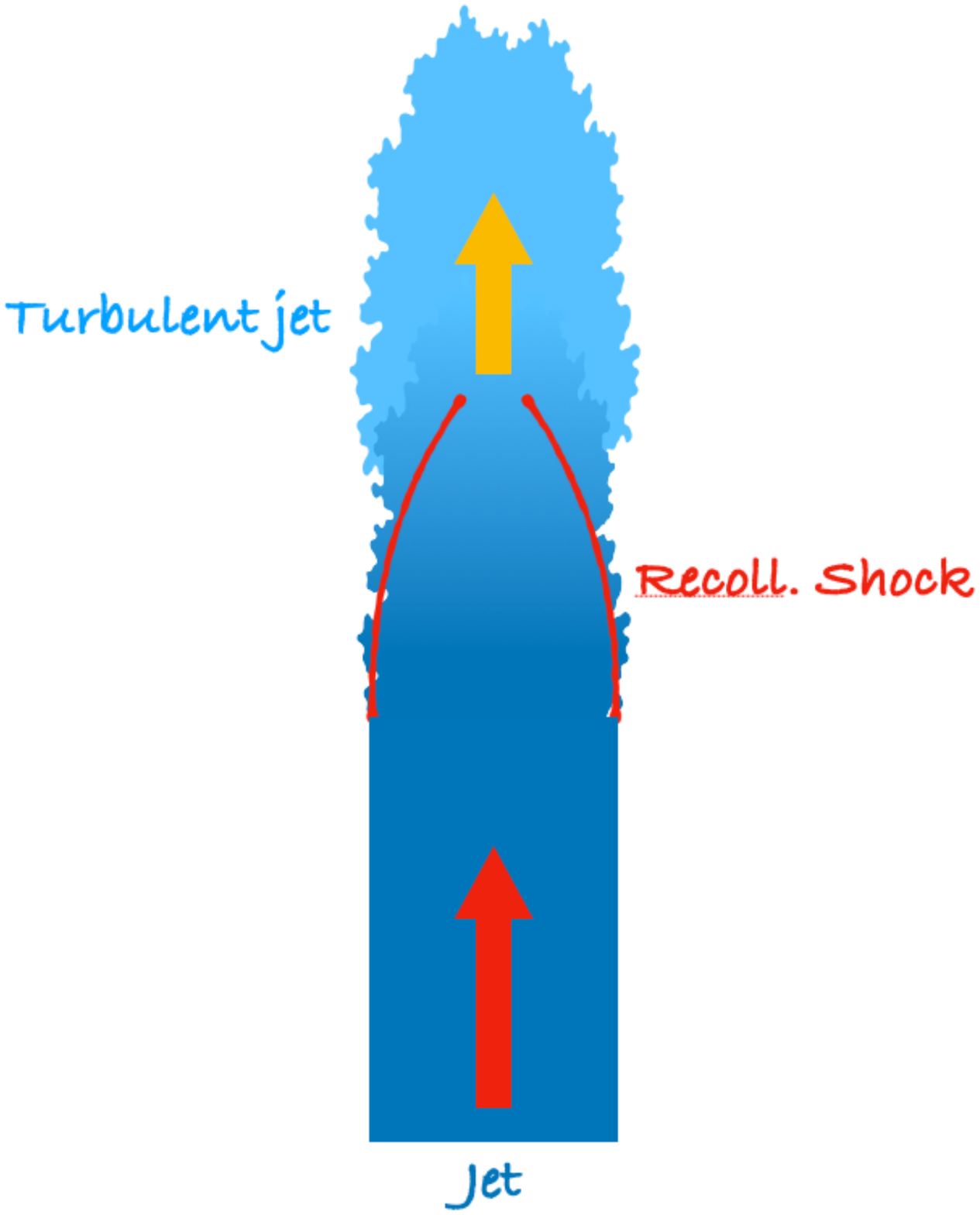}\\
 \vspace*{-1.5truecm}
 \caption{Cartoon of the new scenario proposed for EHBL. Due to the pressure of an external medium the jet  recollimates, triggering the onset of the centrifugal instability recently revealed by 3D simulations. Downstream of the oblique shock, the non-linear development of the instability results in a highly turbulent decelerating flow. Relativistic electrons are accelerated at the shock front and, subsequently, in the turbulent flow through stochastic acceleration, reaching a hard equilibrium distribution.}
 \label{fig:cartoon}
\end{figure}

Along the lines of Zech \& Lemoine (2021) we assume that the jet of EHBL, develops a recollimation shock, induced by the pressure unbalance with the external medium. While Zech \& Lemoine (2021) 
further assume the presence of more than one shock (required to produce an electron distribution hard enough to account for the EHBL spectra), based on the new findings of Gourgouliatos \& Komissarov (2018) and Matsumoto et al. (2021) we propose the scenario depicted in the cartoon (Fig. \ref{fig:cartoon}). The weakly magnetized jet is injected with Lorentz factor $\Gamma$ and recollimates developing a conical shock. The rapid onset of the centrifugal instability leads to the development of macroscopic turbulent eddies and, eventually to the deceleration and the disruption of the jet. We assume that, through the standard cascade process, the turbulent energy injected at large scale is rapidly carried to smaller scales, establishing a standard Kolmogorov spectrum. 

Particles crossing the recollimation shock are accelerated through DSA and, once advected in the downstream region (which flows with bulk Lorentz factor $\Gamma_{\rm d}$),  experience a second, stochastic acceleration stage. In stationary conditions the emission recorded by the observer is produced in the entire downstream region and comprises electrons at different stages of acceleration, from injection (close to the shock) to the equilibrium spectrum (in the far downstream flow).  

\subsection{A simple emission model}

Based on the general scenario developed above we give in the following an example of the spectra resulting by the combination of shock and stochastic acceleration applied to the prototypical EHBL 1ES 0229+200. We remark that we will consider a rather simplified set-up just for illustrative purposes. A more detailed and self-consistent modeling, supported by dedicated numerical simulations, will be reported elsewhere (Costa et al., in prep; Sciaccaluga et al. in prep). Hereafter, all physical quantities (except for the bulk Lorentz factor) are expressed in (downstream) jet rest frame.

The effect of SA is suitably described by a Fokker-Planck equation for the evolution of the energy distribution $N_{\gamma}(t)$ (e.g. Katarzynski et al. 2006, Stawarz \& Petrosian 2008, Tramacere et al. 2011):
\begin{equation}
\begin{split}
\frac{\partial N_{\gamma}(t)}{\partial t}=\frac{\partial}{\partial \gamma}\left [ D_{\gamma} \frac{\partial N_{\gamma}(t)}{\partial \gamma} + \left( \dot{\gamma} - \frac{2D_{\gamma}}{\gamma}\right)N_{\gamma}(t)\right] \\
+ \frac{N_{\gamma}(t)}{t_{\rm esc,\gamma}} + Q_{\gamma},
\label{eq:ngamma}
\end{split}
\end{equation}
where $D_{\gamma}$ is the energy diffusion coefficient, $\dot{\gamma}=\dot{\gamma}_{\rm s} + \dot{\gamma}_{\rm IC}$ is the radiative cooling rate (including synchrotron and IC emission), $t_{\rm esc,\gamma}$ is the escape time and $Q_{\gamma}$ is an injection term. For simplicity we neglect the adiabatic cooling of the particles.

Assuming that SA proceeds through the resonant scattering of MHD waves, with energy spectral density $W(k)$ (with $k$ the wavenumber), the diffusion coefficient can be written as (e.g. Eilek 1979, Kakuwa 2016):
\begin{equation}
D_{\gamma} \simeq \frac{\gamma^2}{t_{\rm acc, \gamma}}\simeq \frac{\gamma ^2\beta_{\rm w}^2 c}{r_{\rm g, \gamma}}\xi(k_{\rm res})
\label{eq:diffcoeff}
\end{equation}
where $r_{\rm g, \gamma}$ is the gyration radius of electrons with Lorentz factor $\gamma$, $k_{\rm res}\approx 1/r_{\rm g, \gamma}$ is the wavenumber of waves resonant with particles with Lorentz factor $\gamma$, $\xi(k_{\rm res})=k_{\rm res}W(k_{\rm res})/U_B$ is the relative amplitude of the turbulent magnetic field energy density in the mode with wavenumber $k_{\rm res}$ and $\beta_{\rm w}$ is the ratio between the wave phase speed and the speed of light. We adopt a power-law form for the turbulence spectrum, $W(k)\propto k^{-q}$, extending from $k_0$ to $k_{\rm max}\gg k_{\rm 0}$. The minimum wavenumber is related to the scale $L$ at which turbulence is injected, $k_0=2\pi/L$. The  total energy density of the turbulent field is assumed to be a fraction $\xi<1$ of the total (turbulent plus ordered) magnetic field energy density. For definitness we assume a Kolmogorov slope for the turbulent spectrum, $q=5/3$. 

We assume that $t_{\rm esc,\gamma}$ can be determined by the space diffusion timescale over the typical source size $r$, 
\begin{equation}
t_{\rm esc,\gamma}\simeq \frac{r^2}{2\kappa_{\gamma}},
\end{equation}
where for the spatial diffusion coefficient one can use the standard expression $\kappa_{\gamma}=c\lambda_{\gamma}/3$. In turn, with the assumed turbulent spectrum, the mean free path $\lambda_{\gamma}$ can be evaluated as (e.g. Liu et al. 2017) $\lambda_{\gamma}=r_{g,\gamma}\xi(k_{\rm res})$. If, for some energies, $\lambda_{\gamma}>r$ we set $t_{\rm esc,\gamma}=r/c$.

We numerically solve  Eq. \ref{eq:ngamma} by using the robust implicit method, especially suited for Fokker-Planck equations, of Chang \& Cooper (1970).

For definitness we assume that scatterings are dominated by Alfven waves, and therefore $\beta_{\rm w} \simeq \beta_{\rm A}=B/(4\pi n_p m_p c^2)^{1/2}$, where $n_p$ is the proton number density.
%

We assume that the (assumed strong) shock accelerates particles with a power law energy distribution with slope $p=2$, a minimum Lorentz factor $\gamma_{\rm min}=2\times 10^3$ (Zech \& Lemoine 2021) and an exponential cut-off at $\gamma_{\rm cut}=2\times 10^5$. The normalization of the distribution is left as a free parameter. These particles are then assumed to describe the (constant in time) injection term $Q_{\gamma}$ in Eq.\ref{eq:ngamma}. 

The region where acceleration and emission occur is modeled as a cylinder with radius $r=0.65\times 10^{16}$ cm and length $l=10r$. This geometry is intended to model the post-shock region, where the instability develops and injects turbulence in the flowing plasma. After the distance $l$ we assume that acceleration and emission are substantially quenched (by adiabatic expansion, decay of magnetic field and jet deceleration). For the produced radiation we use the standard Doppler amplification term (see the discussion in Zech \& Lemoine 2021).

Building on previous studies of EHBL emission (e.g. Costamante et al. 2018), we assume a magnetic field of the order of few mG (although we tune the value to reproduce the SED) and a Doppler factor $\delta=27$.  We further set 
a turbulence injection scale $L= r/5$.

With no escape, the system described by Eq. \ref{eq:ngamma} naturally evolves toward an equilibrium state characterized by a Maxwellian-like distribution with a peak at the energy $\gamma_{\rm p}m_ec^2$ where acceleration is balanced by radiative cooling, i.e. $t_{\rm acc,\gamma}=t_{\rm cool,\gamma}$ (e.g. Eilek 1979, Stawarz \& Petrosian 2008, Katarzynski et al. 2006). 
With $t_{\rm acc,\gamma}$ defined by Eq. \ref{eq:diffcoeff} one derives:
\begin{equation}
\gamma_{\rm p}\simeq 3\times 10^7 \beta_{\rm A,-1}^{3/2} \xi_{-1}^{3/4} B_{-3}^{-5/4}.
\end{equation}
If the escape time is smaller than the cooling timescale the peak is somewhat less pronounced and the maximum energy is determined by the condition $t_{\rm acc,\gamma}\sim t_{\rm esc}$. 
Assuming that the emission region is a cylinder of length $l=10r$, we evolve the distribution up to a time $t_{\rm max}=10 r/c$ (where the system has reached a substantial equilibrium state).

\begin{figure}
\includegraphics[width=0.49\textwidth]{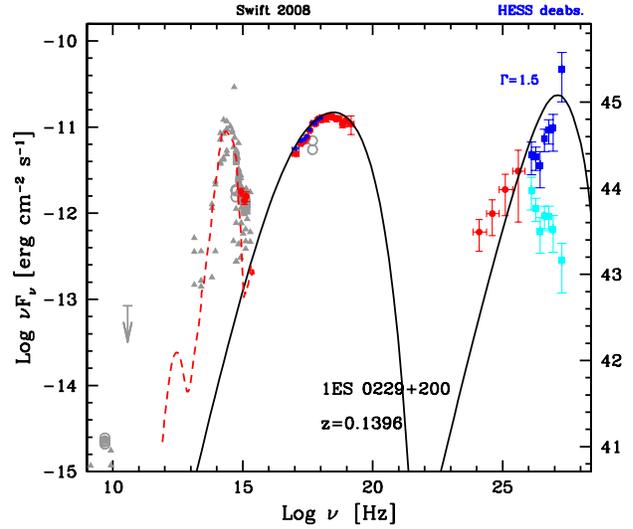}\\
 \vspace*{-2.5truecm}
 \caption{SED of 1ES 0229+200 (data from Costamante et al. 2018) reproduced with the emission model described in the text.}
 \label{fig:sed}
\end{figure}

One realization of the model (obtained by using $\xi=0.2$, $B=3.55$ mG) is reported in Fig.\ref{fig:sed}. With the derived electron density 
the resulting Alfven speed is $\beta_{\rm A}=0.095$ and the (comoving) power associated to the injected electrons is $L^{\prime}_{\rm e}\simeq 5.5\times 10^{39}$ erg s$^{-1}$.

\begin{figure}
\hspace*{-1.2truecm}
\includegraphics[width=0.62\textwidth]{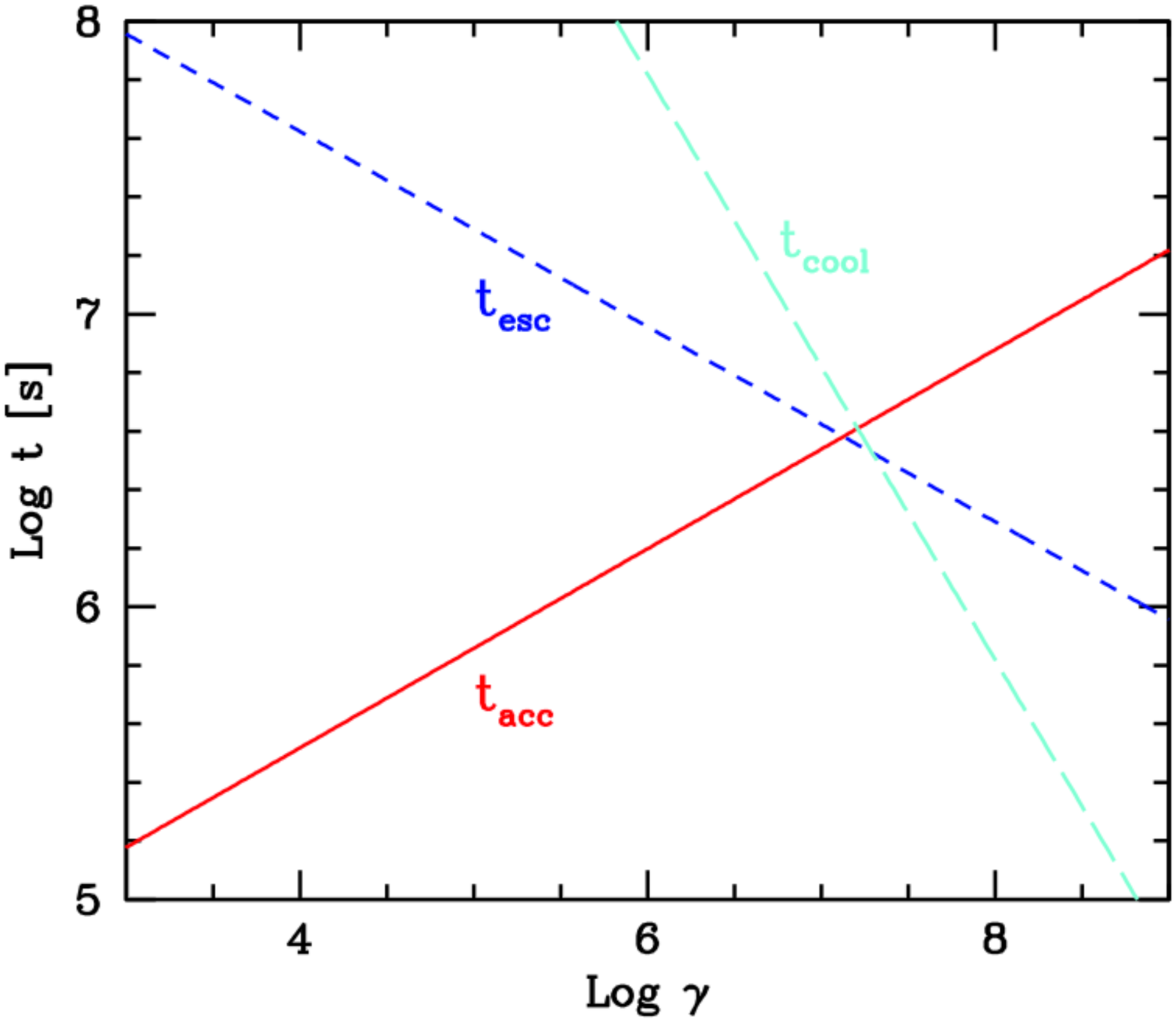}\\
 \caption{Timescales relevant for the stochastic acceleration model described in the text and shown in Fig.\ref{fig:sed}. The lines show the acceleration (red, solid), radiative cooling (light blue, long dashed) and diffusive escape (blue, short dashed) timescales as a function of the electron Lorentz factor.}
 \label{fig:times}
\end{figure}

\begin{figure}
\hspace*{-1.2truecm}
\includegraphics[width=0.63\textwidth]{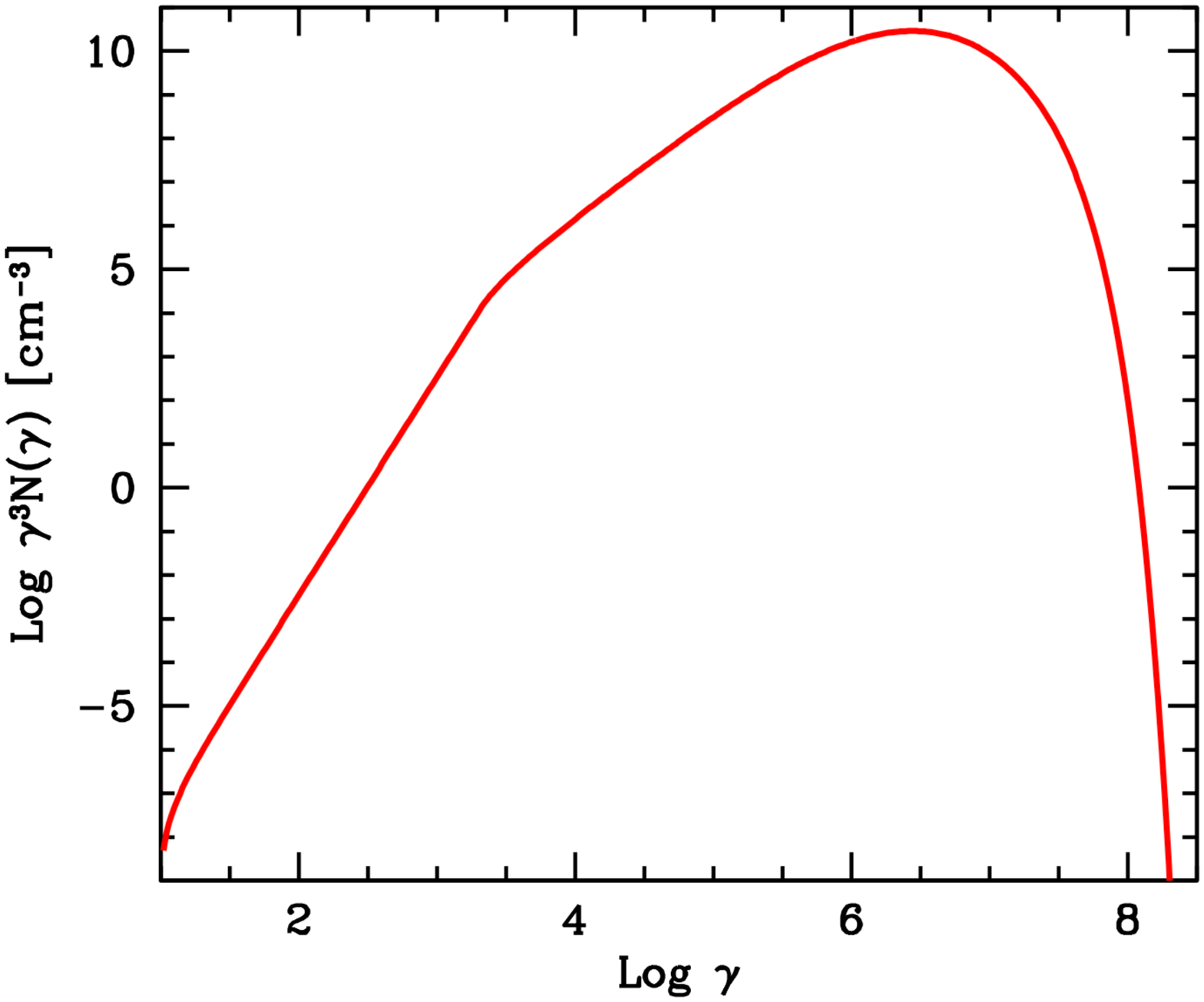}\\
 \vspace*{-.5truecm}
 \caption{The electron energy distribution resulting from SA used to produce the SED in Fig.\ref{fig:sed}.}
 \label{fig:ele}
\end{figure}

The relevant timescales of the system are reported in Fig. \ref{fig:times}. With the parameters adopted in the model of Fig. \ref{fig:sed} the peak of the distribution (located at $\gamma_{\rm p}\approx 10^7$, see Fig. \ref{fig:ele}) is characterized by the near equality of both cooling and escape time.

A few remarks on the simple modeling reported above are in order:
\begin{itemize}
\item While the synchrotron component is quite well reproduced, the slope of the SSC component in the gamma-ray band is harder than the continuum as tracked by the LAT and HESS datapoints; 

\item The system is quite far from equipartition, with the energy density of the relativistic electrons dominating over the magnetic energy density by several orders of magnitudes. In this conditions we expect a strong damping of the turbulence (e.g. Kakuwa 2016), over a timescale $t_{\rm damp}$ much shorter than the cascading time (the time required for turbulent energy to cascade from large to small scales) $t_{\rm casc}$, with $t_{\rm damp}/t_{\rm casc}\sim \beta _{\rm A}^{-1} U_B/U_e\ll 1$. The model is therefore not completely self-consistent, since in this condition the spectrum of the turbulence is modified with respect to the standard power law Kolmogorov spectrum adopted here.
We will present a self-consistent modeling, including the interplay between acceleration and damping, in Sciaccaluga et al. (in prep).

\item We assume a single, homogeneous emission region. The simulations (Matsumoto et al. 2021) suggest that the flow is characterized by a complex structure, with strong turbulent motions. A more realistic modeling should likely include a complex emission region, with a distribution of Doppler factors. A better characterization of the emission region (and turbulence) based on dedicated MHD simulations will be presented elsewhere (Costa et al., in prep).
\end{itemize}

\section{Discussion}

The mechanisms at the base of the peculiar emission displayed by EHBLs are still unclear. The substantially low magnetization inferred from SED modelling points to diffusive shock acceleration as the most likely acceleration process. However, the extreme hardness of the high-energy spectrum is not easily reproduced with the standard relativistic electron spectra expected for DSA. Zech \& Lemoine (2021) suggested that a possible solution is to assume that emitting electrons undergo more than one acceleration stage. This is possible, for instance, in jets undergoing recollimation, which develop a complex structure of multiple shocks, where particles can experience successive acceleration phases, eventually resulting in an electron energy distribution with the required properties.
However, as demonstrated by recent 3D simulations, recollimating weakly magnetized jets are subject to a strong instability which modifies the multiple shock structure which develops in 2D simulations. We have therefore proposed a new framework in which the emission from EHBL has its origin in the combination of acceleration at the recollimation shock followed by stochastic acceleration in the highly turbulent downstream flow. 

In this paper we have explored a highly simplified model based on the hybrid DSA-SA framework outlined above. One of the most important simplifications adopted concerns the energy spectrum of the turbulence, assumed to follow a standard Kolmogorov law. Indeed, as demonstrated by the application to the prototypical 1ES 0229+200, in the specific conditions of EHBL we expect an important interplay between particles and turbulence with the subsequent strong damping of the turbulent waves. The study of a self-consistent model will be reported elesewhere (Sciaccaluga et al. in prep).

An important point of the model concerns the location of the recollimation region (and the related issue of the nature of the material thought to provide the pressure required to reconfine the jet). In the model we assume a relatively small jet radius, of the order of $r\sim 10^{16}$ cm. This size is fixed by the condition that the energy density of the synchrotron radiation is high enough to have the right level of the inverse Compton component, fixed by high-energy data. Indeed, a substantially larger radius would imply a reduced radiation energy density (proportional to $\sim 1/r^2$) that should be compensated by a corresponding reduction of the magnetic energy density. The much lower magnetic field, however, would imply a reduced Alfven speed, thus greatly reducing the efficiency of the SA (the acceleration timescale is $t_{\rm acc} \propto \beta^{-2}_{\rm A}$). Therefore, the size is ultimately constrained by requirement that SA is efficient enough to produce particles emitting in the X-ray band. In our model the radius $r$ is associated to the cross section of the post-reconfinement flow where most of the dissipation occurs, that can be substantially smaller ($\lesssim1/10$) than the jet radius (e.g. Bodo \& Tavecchio 2018). Considering  a standard aperture angle $\theta_{\rm j}\simeq 0.1$ (but there are indications that it can be substantially smaller, e.g. Pushkarev et al. 2009) we can thus locate the recollimation region at the pc scale. At these distances the jet could be confined by gas expelled by the central regions in the form of winds or outflows (e.g. Globus \& Levinson 2016).

As demonstrated by the simulations reported by Matsumoto et al. (2021), the presence of magnetic fields can effectively suppress the recollimation instability.  Although the precise value of the critical magnetization is likely dependent on the details of the system (magnetic field geometry, pressure contrast, profile of external pressure), present results suggest that for modest magnetic field the jet is stabilized (the simulations indicate a threshold value of the magnetization $\sigma\sim 0.06$) . In this framework it is therefore tempting to speculate that the special properties characterizing EHBL are related to the very low magnetization of their jets. Only for these sources the jet develops a highly turbulent downstream flow suitable for SA to operate. For other classes of BL Lacs (and blazars in general), the larger magnetic field will ensure a stable jet, thus leaving shock acceleration as the only operating acceleration mechanism. 

It is worth noticing that the rather small magnetization inferred for EHBL (but also for less extreme BL Lacs, see e.g. Tavecchio \& Ghisellini 2016) is quite puzzling in the standard picture for jet acceleration, which envisions an initially highly magnetized outflow, progressively accelerating through the conversion of magnetic to kinetic flux, until equipartition is established (e.g. Komissarov et al. 2007). A possibility to reconcile the low magnetization with the magnetic acceleration scenario is to assume a phase of highly efficient dissipation of magnetic fields through magnetic reconnection (Giannios \& Uzdensky 2019, Zhang \& Giannios 2021), although the distance at which the complete conversion should occur (tens of parsec) is larger than those commonly inferred for the "blazar zone".

A last observations concerns the polarization properties expected for the emission from EHBL in this scenario. Since most of the emission comes from particles emitting in a highly turbulent magnetic field, we expect that the synchrotron emission (responsible for the radiation up to the X-ray band) displays a very low degree of polarization. Polarimetric measurements in the optical band for EHBL are hampered by the weakness of the jet emission in this band, strongly dominated by the unpolarized emission of the host elliptical galaxy (see Fig. \ref{fig:sed}). The now operative {\it IXPE} satellite (Weisskopf et al. 2016) will measure for the first time the polarimetric properties of the X-ray emission of blazars. A quite robust prediction of our scenario is therefore that for EHBL {\it IXPE} will measure a very low degree of polarization.

\section*{Acknowledgments}
We thank Gianluigi Bodo, Andreas Zech and Martin Lemoine for useful discussions. We thank the referee for constructive comments. FT acknowledges contribution from the grant INAF Main Stream project "High-energy extragalactic astrophysics: toward the Cherenkov Telescope Array".

\section*{Data availability}

Data available on request.

\end{document}